\documentclass[12pt]{article}
\usepackage{a4wide,graphicx,amsmath,amssymb,hyperref}
\usepackage[numbers,sort&compress]{natbib}
\usepackage{hypernat}
\usepackage{subfigure}
\usepackage{multirow}
\def\slash#1{#1 \hskip-0.45em /}
\def\beq{\begin{equation}}
\def\eeq{\end{equation}}
\def\bea{\begin{eqnarray}}
\def\eea{\end{eqnarray}}

\newcommand{\newc}{\newcommand}
\newc{\pt}{p_T}
\newc{\MW}{M_W}

\newc{\Wp}{W^+}
\newc{\Wm}{W^-}
\newc{\Wpm}{W^{\pm}}
\newc{\ttbar}{t\bar{t}}
\newc{\HT}{H_T}

\begin{document}
  \titlepage
  \begin{flushright}
    Cavendish-HEP-2010/06 \\
    April 2010 \\
  \end{flushright}
  \vspace*{0.5cm}
  \begin{center}
    {\Large \bf Charge asymmetry in W + jets production at the LHC}\\
    \vspace*{1cm}
    \textsc{C.H.~Kom and W.J. Stirling}\\
    \vspace*{0.5cm}
         Cavendish Laboratory, University of Cambridge, CB3 0HE, UK\\
  \end{center}
  \vspace*{0.5cm}
  \begin{abstract}
    The charge asymmetry in $W^\pm +$~jets production at the LHC can
    serve to calibrate the presence of New Physics contributions. We
    study the ratio $\sigma(W^+ + n\; \mbox{jets})/\sigma(W^-+ n\;
    \mbox{jets})$ in the Standard Model for $n \leq 4$, paying
    particular attention to the uncertainty in the prediction from
    higher-order perturbative corrections and uncertainties in parton
    distribution functions. We show that these uncertainties are
    generally of order a few percent, making the experimental
    measurement of the charge asymmetry ratio a particularly useful
    diagnostic tool for New Physics contributions.
  \end{abstract}


\section{Introduction}
At $p \bar p$ colliders such as the Fermilab Tevatron, $W^+$ and $W^-$
bosons are produced in equal quantities, i.e. $\sigma(W^+) =
\sigma(W^-)$. In contrast, at the CERN LHC $pp$ collider, $\sigma(W^+)
\approx 1.3\,\sigma(W^-)$. This charge asymmetry is directly related
to the dominance of $u$ quarks to $d$ quarks in the proton,
$R_{ud}(x,Q^2) = u(x,Q^2)/d(x,Q^2) > 1$. In standard parton
distribution function (PDF) global fits, $R_{ud} \approx 1$ for $x \ll
1$ and increases monotonically as $x$ increases. The charge asymmetry
ratio $\sigma(W^+)/\sigma(W^-) \neq 1$ is a feature of both the
inclusive $W^\pm$ total cross sections, and also of more exclusive
$W^\pm+n$~jet cross sections. An important feature of this ratio is
that it is theoretically a very stable quantity.  In particular, it is
expected to be stable with respect to electroweak parameter values and
higher-order (electroweak and QCD) perturbative corrections, because
the couplings and kinematics of the $\Wp$ and $\Wm$ subprocesses are
essentially the same.

Precise measurements of the $W$ charge asymmetry at the LHC can
therefore yield further information on the $u/d$ parton ratio, see for
example Ref.~\cite{{Martin09:2009iq}}. Here we look at a different
aspect of the asymmetry, namely that we can use the very precise
knowledge of the $u/d$ ratio to calibrate the Standard Model (SM)
$W+n$~jet background to New Physics (NP) at the LHC, since typically
$\sigma_{\rm NP}(X \to W^+ + {\rm jets}) = \sigma_{\rm NP}(X \to W^- +
{\rm jets})$.\footnote{A discussion and preliminary results were
  presented in Ref.~\cite{Stirling:2009}.  }  In fact within the SM,
there are a number of interesting physics processes, for example
$\ttbar$ and Higgs boson production,
\begin{eqnarray*}
  \begin{tabular}{ccccccc}
    $gg$&$\to$&$t \bar t$&$\to$&$W^+W^- b \bar b$&$\to$&$W^\pm(\to l^\pm + \nu) + 4\; {\rm jets}$, \\
    $gg$&$\to$&$H$       &$\to$&$W^+W^-$         &$\to$&$W^\pm(\to l^\pm + \nu) + 2\; {\rm jets}$,
  \end{tabular}
\end{eqnarray*}
that give rise to equal numbers of $W^+$ and $W^-$ bosons.  NP
examples can be found for instance in gluino pair production in
supersymmetry, where cascade decays of the gluino pair into one
charged lepton plus missing energy and jets are expected to have the
same cross sections for opposite lepton charges.  The fundamental idea
is that any observed deviation from the predicted SM value of the
ratio \beq R^\pm(n)= \frac{\sigma(W^+ + n\; {\rm jets})}{\sigma(W^- +
  n\; {\rm jets})} \eeq could signal the presence of a NP contribution
in the $W +$~jets sample.\footnote{The $W^\pm$ are in practice always
  assumed to decay to a single generation of leptons.} In particular,
if $ \frac{1}{2}\sigma_{\rm NP} \equiv \sigma_{\rm NP}(X \to W^++{\rm
  jets})=\sigma_{\rm NP}(X \to W^-+{\rm jets})$, and $\sigma_{\rm SM}
\equiv \sigma_{\rm SM}(X \to W^++{\rm jets}) + \sigma_{\rm SM}(X \to
W^- + {\rm jets})$, then \beq f_{\rm NP} = \frac{2(R^\pm_{\rm
    SM}-R^\pm_{\rm exp.})}{(R^\pm_{\rm SM} + 1) (R^\pm_{\rm exp.}-1)},
\eeq where $f_{\rm NP} = \sigma_{\rm NP} / \sigma_{\rm SM}$ is the
ratio of the NP and SM cross sections, and $R^\pm_{\rm exp.}$ and
$R^\pm_{\rm SM}$ are the experimentally measured and SM expectation of
the cross-section ratio $R^\pm$ respectively.  Hence a measurement of
$R^\pm(n)$, combined with the SM theoretical prediction, enables a
value of $f_{\rm NP}$ to be extracted. To give a very simple numerical
example, the contribution of $\ttbar \to W + 4\; {\rm jet}$
production to the SM $W + 4\; {\rm jet}$ final state (at the 14~TeV
LHC and with typical parameters and cuts defined in
Eq.~(\ref{eq:cutsdef}) below) reduces $R^\pm(4)$ from 1.55 to 1.22.
Hence if the uncertainty on the SM prediction of 1.55 is small enough,
the presence of $t\bar t$ contributions can be detected.  A
preliminary study of using the $W^\pm$ charge asymmetry to help
identify the top quark signal in early LHC running has already been
reported by the CMS collaboration~\cite{CMS}.

There are basically only two significant sources of theoretical
uncertainty on $R^\pm(n)$: unknown higher-order pQCD corrections to
the subprocess cross sections and PDF uncertainties.  We already know
from studies~\cite{{Martin09:2009iq}} of the {\em total} $W^\pm$ cross
sections (see Table~\ref{tab:w+w-totalrat} below) that both
uncertainties are small. Here we consider the corresponding
uncertainties on the $W^\pm +$~jets cross sections. In the following,
we will first describe our calculational procedure, and then present
results for up to and including $W+4$~jet production at the LHC.  We
will also demonstrate a connection between the exact calculation and
the high-energy (`BFKL') approximation which may allow $R^{\pm}(n\geq
5)$ to be estimated.


\section{Calculational framework and results}

We begin by reviewing the results for the charge asymmetry ratio for
total cross sections, see Table~\ref{tab:w+w-totalrat}. These have
already been discussed in some detail in Ref.~\cite{Martin09:2009iq}
where a similar table was presented. Note that the ratio decreases
slightly when going from leading order (LO) to next-to-leading order
(NLO), and then appears to be perturbatively stable. This is more a
reflection of the differences between the LO, NLO and
next-to-next-to-leading order (NNLO) PDFs than the impact of the
higher-order subprocess corrections. For example, the 14~TeV LHC LO
ratio $R^\pm=1.365$ reduces to $R^\pm=1.312$ when MSTW2008 NLO partons
are used in the leading-order cross-section calculations, and then
increases slightly to $R^\pm=1.325$ when the explicit NLO corrections
are included.\footnote{Real gluon emission at ${\cal O}(\alpha_S)$
  causes the typical parton $x$ values to increase slightly compared
  to the leading-order calculation, which in turn gives rise to a
  larger $u/d$ ratio and hence a slightly larger $R^\pm$.}

\begin{table}[!ht]
  \centering
  \begin{tabular}{|l|c|c|}
    \hline\hline
       &  $\sqrt{s} = 7$~TeV & $\sqrt{s} = 14$~TeV\\
    \hline
    MSTW 2008 LO   &  $1.463\pm 0.014$ & $1.365\pm 0.011$ \\
    MSTW 2008 NLO  &  $1.422\pm 0.012$ & $1.325\pm 0.010$ \\
    MSTW 2008 NNLO &  $1.429\pm 0.013$ & $1.328\pm 0.011$ \\
    \hline\hline
  \end{tabular}
  \caption{Predictions for the ratio of $W^+$ and $W^-$ total cross
    sections at the LHC at LO, NLO and NNLO pQCD, including the
    one-sigma (68\% cl) PDF uncertainties, with $\mu_R=\mu_F=\MW$.}
  \label{tab:w+w-totalrat}
\end{table}

We note also that the ratio of $W^+$ and $W^-$ cross sections is
strongly dependent on the $W$ boson rapidity $y$. Indeed at large $y$
we have \beq R^\pm \sim \frac{u(x,\MW^2)}{d(x,\MW^2)}, \quad
\mbox{with}\ x = \frac{\MW}{\sqrt{s}}\; \exp (y) , \eeq so that, at
least for the MSTW2008 PDFs, $R^\pm \to \infty$ as $y\to y_{\rm max}$.
The ratio also decreases with increasing collider energy $\sqrt{s}$,
as smaller $x$ values are probed.

For $W+n$~jet production we use MSTW2008 PDFs \cite{Martin09:2009iq}
throughout and calculate the cross sections using the MCFM package
\cite{MCFM} for $n=0,1,2$ (LO and NLO) and $n=3$ (LO only). For the
$n=4$ LO calculation we use an adaptation of the VECBOS package
\cite{VECBOS}.  The electroweak parameters used are the same as
\cite{Martin09:2009iq}.  We note that the NLO corrections to the $n=3$
cross section have recently been calculated \cite{Berger:2009zg}: we
will comment on their likely impact on our results below. We follow
the standard practice of setting the renormalisation and factorisation
scales equal, i.e. $\mu_R=\mu_F=\mu$, in all calculations.

In order to define realistic $W+n$~jet cross sections we need to specify
cuts on the final-state particles. We use a set of standard
cuts:
\bea
\begin{tabular}{lll}
$\vert\eta_l\vert^{\rm max} = 2.5$,& $p_{Tl}^{\rm min} = 20\; {\rm GeV}$,& $\slash{E}_{T}^{\rm min} = 20\; {\rm GeV}$ , \\
$\vert\eta_j\vert^{\rm max} = 2.5$,& $p_{Tj}^{\rm min} = 40\; {\rm GeV}$,& $\Delta R_{jj} > 0.7$ ,
\end{tabular}
\label{eq:cutsdef}
\eea
typical of the LHC general purpose detectors. We will also study how
$R^\pm(n)$ varies with $p_{Tj}^{\rm min} $, since many NP signals are
expected to produce energetic jets in the final state.

\begin{figure}[ht]
\begin{center}
\includegraphics[width=0.53\textwidth]{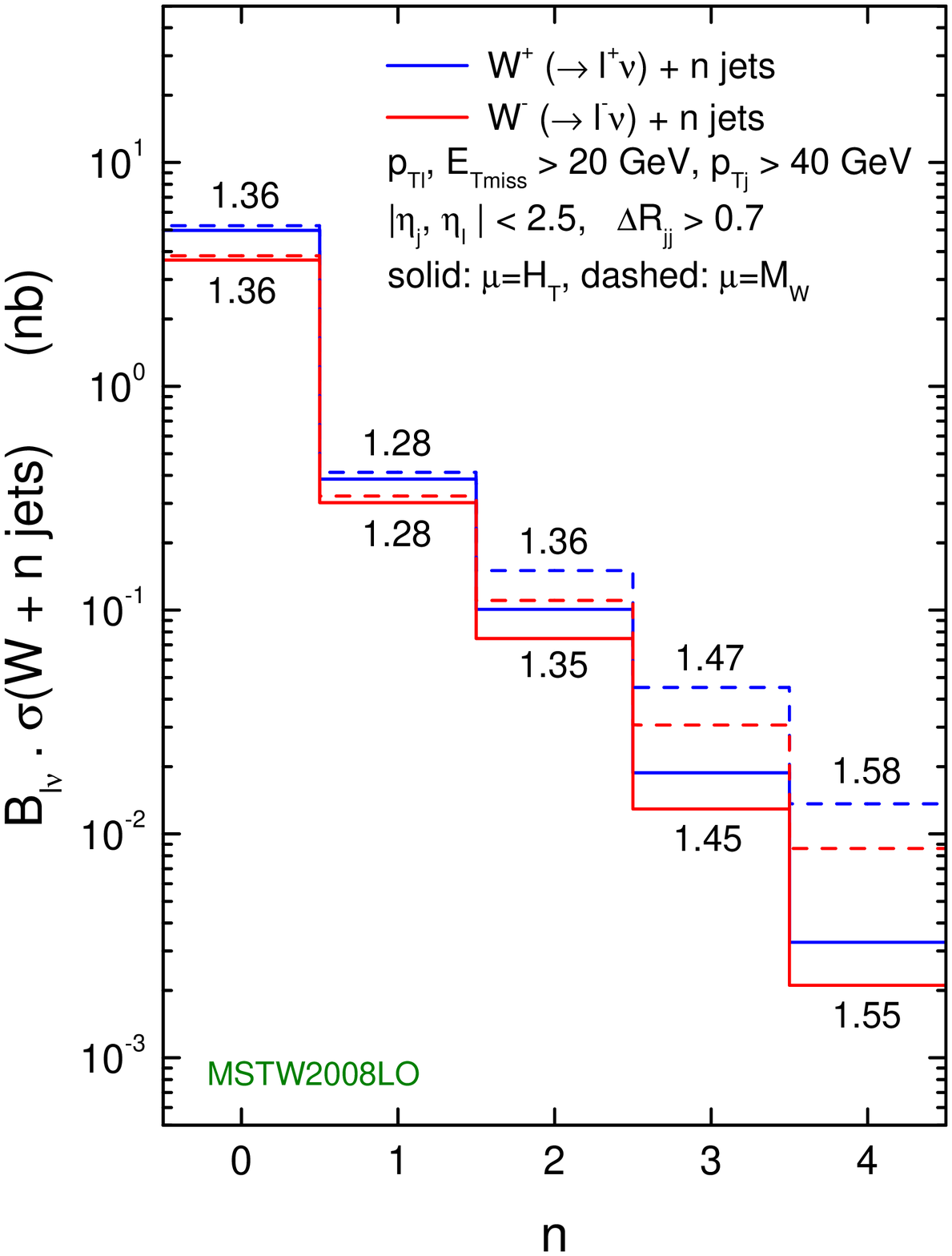}
\caption{Predicted cross sections times leptonic branching ratio for
  $W^\pm +n$~jet production ($n=0,1,2,3,4$) at the 14~TeV LHC, using
  the standard set of cuts defined in Eq.~(\ref{eq:cutsdef}) and
  calculated at leading order with scales $\mu\,(= \mu_R = \mu_F)=M_W$
  and $\mu = \HT$, where $\HT$ is the scalar sum of the transverse
  momenta of the final-state jets and leptons, $\HT = \sum_i
  p_{Ti}$. The cross-section ratios $R^\pm(n)$ are marked on the
  plot. }\label{fig:xsecRatio}
\end{center}
\end{figure}

The leading-order cross sections, evaluated at two different scales
$\mu=M_W$ and $\mu = \HT$ (the scalar sum of transverse momenta of all
visible particles $\sum_i p_{Ti}$), together with their corresponding
$R^\pm(n)$ values, are shown in Fig.~\ref{fig:xsecRatio} for the
14~TeV LHC.  Note that variation with scale choice of the ratios is
much smaller than the variation in the absolute values of the cross
section. With these cuts and scale choices, the charge asymmetry
ratios lie roughly in the range $[1.25,1.60]$ with a noticeable
dependence on $n$.

\begin{table}[ht]
  \centering
  \begin{tabular}{|ll|c|c|c|c|c|}
    \hline\hline
    &&&&&& \\
    & $n$    & $\mu = \MW$  &   $\mu = E^W_T$  & $\mu = \HT/2$  & $\mu = \HT$  & $\mu = 2\HT$ \\
    &&&&&& \\
    \hline
    \multirow{5}{*}{LO}
    & 0  & 1.365(1) & 1.365(1) & 1.364(1) & 1.359(1) & 1.355(1)\\
    & 1  & 1.276(1) & 1.275(1) & 1.275(1) & 1.276(1) & 1.276(1)\\
    & 2  & 1.358(1) & 1.351(1) & 1.349(1) & 1.351(1) & 1.352(1)\\
    & 3  & 1.472(2) & 1.455(2) & 1.446(2) & 1.451(2) & 1.454(2)\\
    & 4  & 1.58(1)  & 1.55(1)  & 1.54(1)  & 1.55(1)  & 1.54(1) \\
    \hline
    \multirow{3}{*}{NLO} 
    & 0  & 1.310(1) & 1.310(1) & 1.309(1) & 1.309(1) & 1.309(1)\\
    & 1  & 1.270(1) & 1.268(1) & 1.274(1) & 1.269(1) & 1.265(1)\\
    & 2  & 1.326(5) & 1.328(4) & 1.341(2) & 1.333(2) & 1.335(2)\\
    \hline\hline
  \end{tabular}
  \caption{The renormalisation and factorisation scale dependence
    ($\mu =\mu_R=\mu_F$) of the leading-order (and next-to-leading
    order for $n=0,1,2$) ratio $R^\pm(n)$, for the standard set of
    cuts defined in Eq.~(\ref{eq:cutsdef}) at the 14~TeV LHC. Here
    $E^W_T$ is the transverse energy of the $W$ boson,
    $E^W_T=\sqrt{\MW^2 + p_{T}(W)^2}$, and $\HT$ is the scalar sum of
    the transverse momenta of the final-state jets and leptons, $\HT =
    \sum_i p_{Ti}$. The numbers in brackets are the estimated
    calculational errors on the final displayed significant figure.}
  \label{tab:scalevar}
\end{table}

\begin{figure}[!ht]
  \begin{center}
    \begin{tabular}{cc}
      \subfigure[$W+0$ jets]{
        \scalebox{0.75}{
          \includegraphics{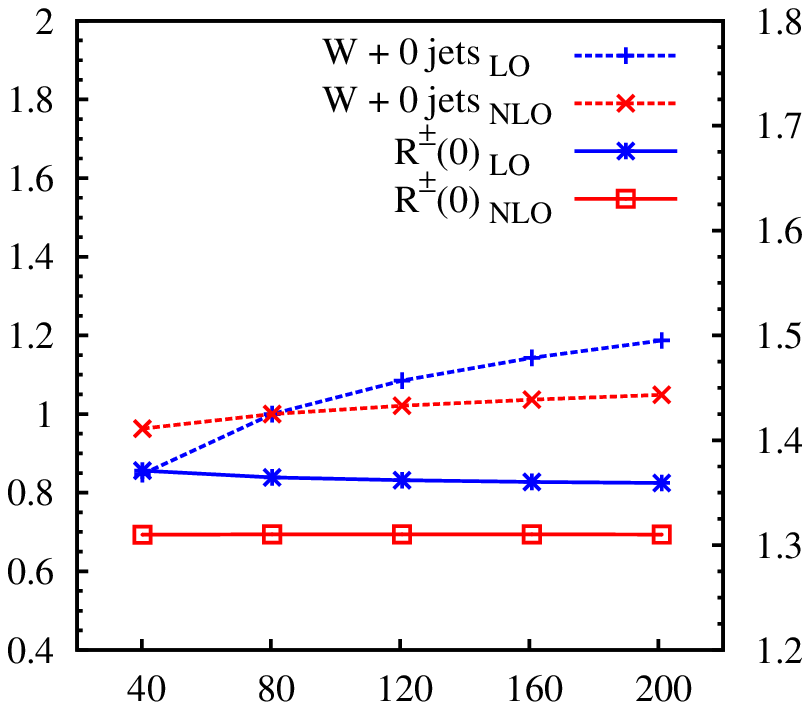}
          \label{fig:xsec_norm_W0j}
        }
        \put(-210,65){\rotatebox{90}{$\frac{\sigma(\mu)}{\sigma(\mu=\MW)}$}}
        \put(-25,162){\rotatebox{0}{$R^\pm(n)$}}
        \put(-115,-10){$\mu_F$ (GeV)}
        \put(-80,-20){$ $}
      }
      &
      \subfigure[$W+1$ jet]{
        \scalebox{0.75}{
          \includegraphics{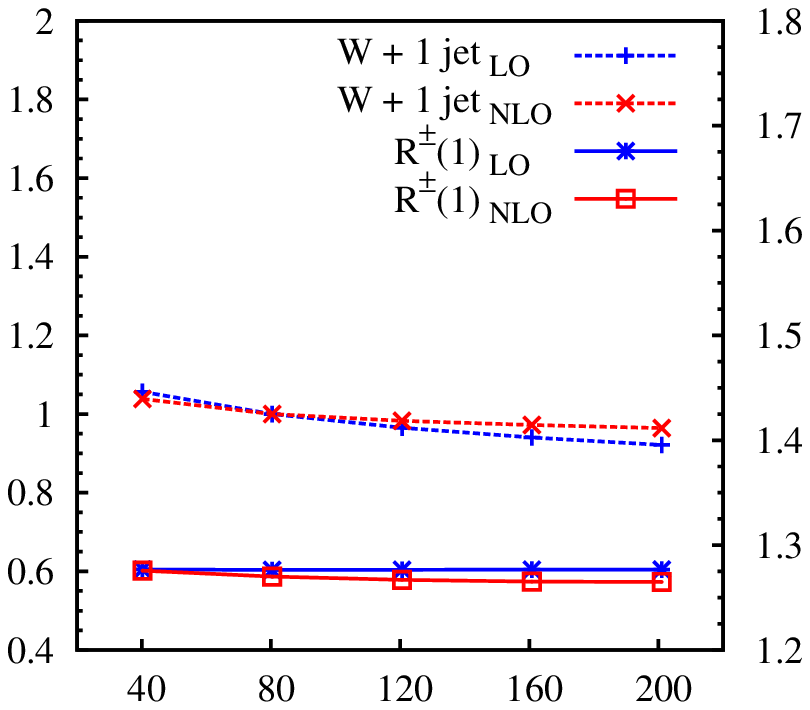}
          \label{fig:xsec_norm_W1j}
        }
        \put(-210,65){\rotatebox{90}{$\frac{\sigma(\mu)}{\sigma(\mu=\MW)}$}}
        \put(-25,162){\rotatebox{0}{$R^\pm(n)$}}
        \put(-115,-10){$\mu$ (GeV)}
        \put(-80,-20){$ $}
      }
      \\
      \subfigure[$W+2$ jets]{
        \scalebox{0.75}{
          \includegraphics{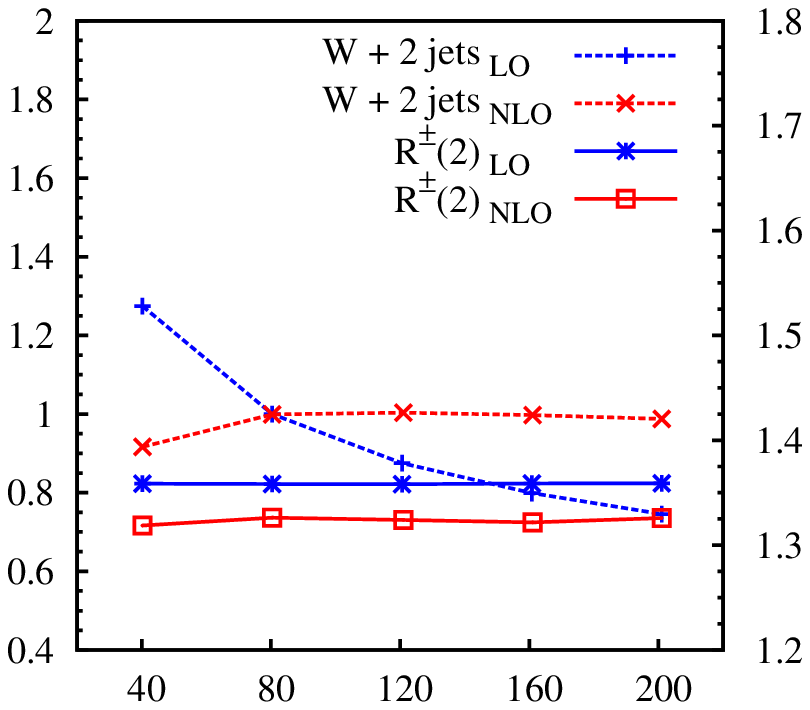}
          \label{fig:xsec_norm_W2j}
        }
        \put(-210,65){\rotatebox{90}{$\frac{\sigma(\mu)}{\sigma(\mu=\MW)}$}}
        \put(-25,162){\rotatebox{0}{$R^\pm(n)$}}
        \put(-115,-10){$\mu$ (GeV)}
        \put(-80,-20){$ $}
      }
      &
      \subfigure[$W+3,4$ jets (LO only)]{
        \scalebox{0.75}{
          \includegraphics{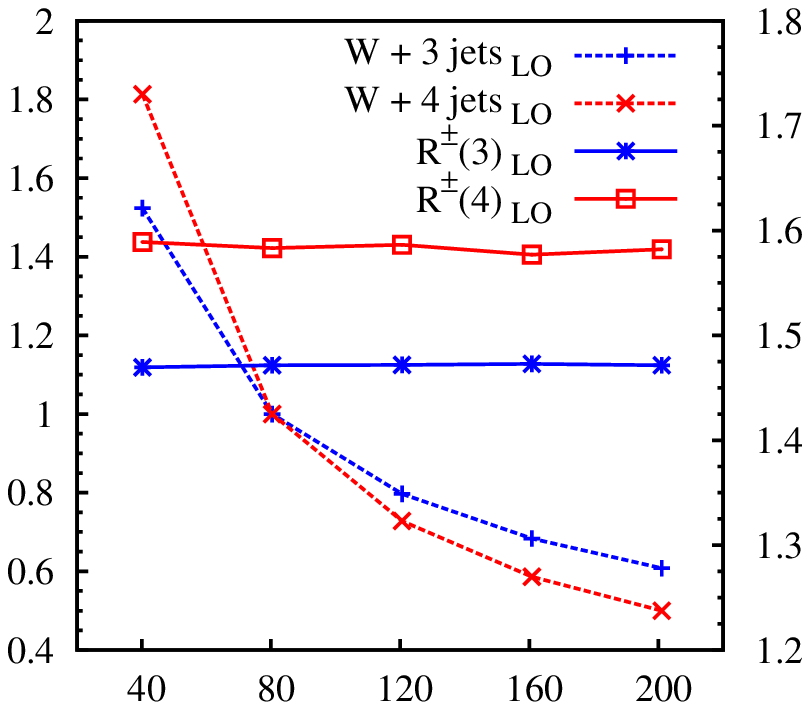}
          \label{fig:xsec_norm_W3j}
        }
        \put(-210,65){\rotatebox{90}{$\frac{\sigma(\mu)}{\sigma(\mu=\MW)}$}}
        \put(-25,162){\rotatebox{0}{$R^\pm(n)$}}
        \put(-115,-10){$\mu$ (GeV)}
        \put(-80,-20){$ $}
      }
      \\
    \end{tabular}
    \caption{Ratios of cross sections for $W^\pm +n$~jet production
      ($n=0,1,2,3,4$) to the values obtained with scale $\mu = \mu_R =
      \mu_F=\MW$ as a function of $\mu$.  For $n=0,\; 1$ and $2$, the
      cross sections are calculated at both leading and
      next-to-leading order.  The ratios $R^\pm(n)$ are also shown
      (right-hand axis).}
    \label{fig:scalevar}
  \end{center}
\end{figure}

To study the renormalisation and factorisation scale dependence of the
$R^\pm(n)$ predictions in more detail, we choose a variety of scales
$\mu$, from fixed scales ($\propto M_W$) to `dynamical' scales that
depend on the lepton and jet kinematics.  The results for $R^\pm(n)$
at leading order and next-to-leading order (where available in MCFM),
with standard cuts (\ref{eq:cutsdef}) at the 14~TeV LHC, are given in
Table~\ref{tab:scalevar}, using scales $\mu=\MW$, the transverse
energy of the $W$ ($E^W_T=\sqrt{\MW^2 + p_T(W)^2}$) and multiples of
$\HT$.  We also show the variation of individual cross sections and
$R^\pm(n)$ as a function of fixed scale $\mu$ in
Fig.~\ref{fig:scalevar}.  As expected, the NLO cross sections show
much weaker scale dependence compared with the LO results.  We also
note the slight modification of the cross-section ratios when going
from leading to next-to-leading order, an effect already noticed in
the total cross sections.  However, it is apparent that even at
leading order the cross-section ratios are remarkably stable with
respect to variation of scale choice.

\begin{figure}[ht]
  \begin{center}
    \includegraphics[width=0.55\textwidth]{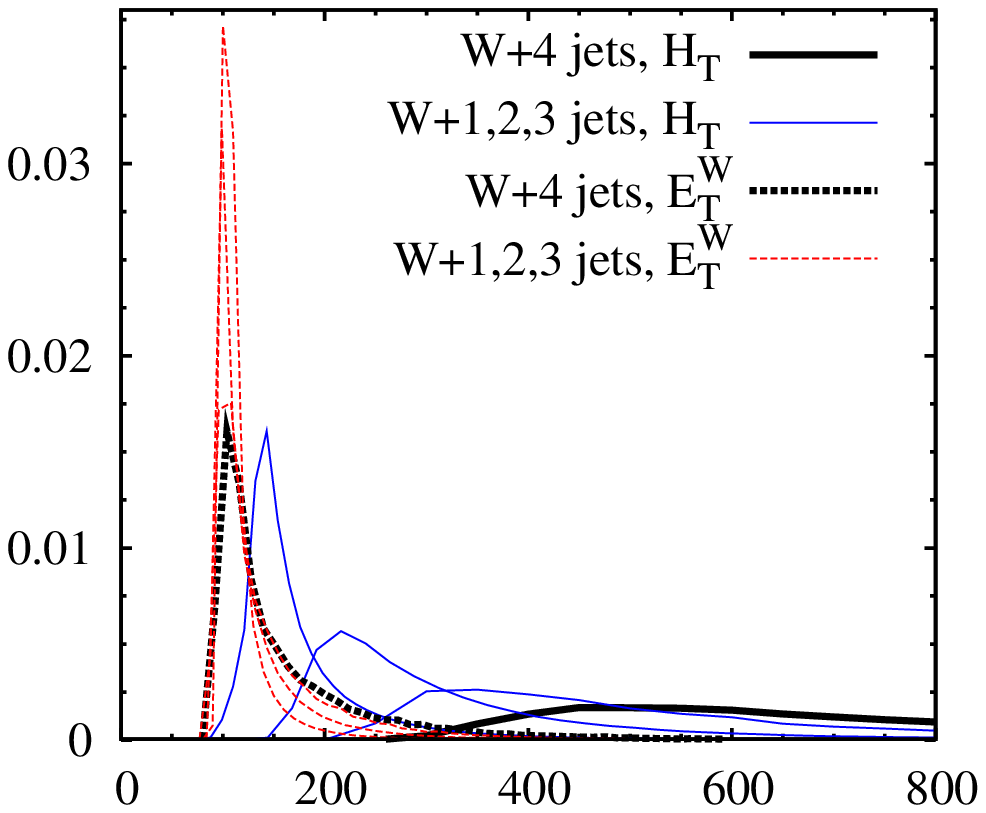}
    \put(-275,40){\rotatebox{90}{Normalised distribution}}
    \put(-135,-10){GeV}
    \put(-80,-15){$ $}
    \caption{Normalised distributions of the $W$ transverse energy
      $E^W_T$ (dashed) and transverse momenta scalar sum $\HT$ (solid)
      for $W^\pm+n$~jet production ($n=1,2,3,4$) at the 14~TeV LHC,
      using the standard set of cuts defined in Eq.~(\ref{eq:cutsdef})
      and calculated at leading order with scales $\mu\,(= \mu_R =
      \mu_F)=M_W$. As $n$ increases, both distributions move towards
      higher values.  The $n=4$ distributions are displayed in
      black.}\label{fig:dist_HT_WET}
  \end{center}
\end{figure}

As the number of jets increases, new kinematic configurations open up,
leading to a broader range of internal energy and momentum scales.  To
illustrate this, we show in Fig.~\ref{fig:dist_HT_WET} the
distribution of two dynamical quantities, $\HT$ and $E^W_T$, for
different number of jets $n$.  The increase in the spread of these two
variables with $n$ lends support to the use of dynamical rather than
fixed scales $\mu$. Indeed in Ref.~\cite{Berger:2009zg}, the use of
$\mu=\HT$ was advocated for $W+$~jets production, since with this
scale choice many representative kinematic distributions were shown to
be positive with stable K--factors.  We therefore choose $\HT$ as our
default scale choice when predicting $R^\pm(n)$.  The resulting
ratios, as well as the individual cross sections, are displayed in
Table~\ref{tab:xsecCut} for both 7 and 14~TeV LHC.  The $\ttbar$ cross
sections are also shown for comparison.

\begin{table}[!ht]
  \centering
  \begin{tabular}{|c|c|cc|c||cc|c|}
    \hline\hline
    \multicolumn{2}{|c|}{}&\multicolumn{3}{c||}{7 TeV}&\multicolumn{3}{c|}{14 TeV}\\
    \cline{3-8}
    \multicolumn{2}{|r|}{n}&$\sigma^+(n)$&$\sigma^-(n)$&$R^\pm(n)$&$\sigma^+(n)$&$\sigma^-(n)$&$R^\pm(n)$\\
    \hline
    \multirow{5}{*}{LO}
    &0 & 2860(1)  & 1800(1)  & 1.589(1) & 4980(1)  & 3665(1)  & 1.359(1)\\
    &1 & 162.8(1) & 110.2(1) & 1.478(1) & 385.4(1) & 302.1(1) & 1.276(1)\\
    &2 & 35.68(2) & 22.53(1) & 1.584(1) & 100.9(1) & 74.74(5) & 1.351(1)\\
    &3 & 5.339(4) & 3.099(2) & 1.723(2) & 18.78(2) & 12.94(1) & 1.451(2)\\
    &4 & 0.734(3) & 0.392(1) & 1.87(1)  & 3.28(2)  & 2.11(1)  & 1.55(1) \\
    \hline
    \multirow{3}{*}{NLO}
    &0 & 3210(1)  & 2114(1)  & 1.519(1) & 5234(2)  & 3997(1)  & 1.309(1)\\
    &1 & 211.1(1) & 145.2(1) & 1.454(1) & 478.3(2) & 376.9(1) & 1.269(1)\\
    &2 & 42.53(4) & 27.31(2) & 1.557(2) & 114.3(1) & 85.69(8) & 1.333(2)\\
    \hline\hline
    \multicolumn{2}{|r|}{$\ttbar$}&0.6473(8)&0.6473(8)&1&3.297(5)&3.297(5)&1\\
    \hline\hline
  \end{tabular}
  \caption{Predicted cross sections (in pb) for $\sigma^{\pm}(n)\equiv
    \sigma(W^\pm (\to e^{\pm}\nu) +n~{\rm jets})$ production at 7 and
    14~TeV at the LHC.  Both the renormalisation ($\mu_R$) and
    factorisation ($\mu_F$) scales are set equal to $\HT$, the scalar
    sum of the transverse momenta of jets and leptons.  In obtaining
    these cross sections, the standard set of cuts defined in
    Eq.~(\ref{eq:cutsdef}) is applied.  For comparison, the
    leading-order cross section $\sigma(\ttbar \to e^{\pm}\nu +4j)$ is
    also shown.  In this case, the scale $\mu = \mu_R =
    \mu_F=m_t=171.3$~GeV is used.  The numbers in brackets are the
    estimated calculational errors on the final displayed significant
    figure.}\label{tab:xsecCut}
\end{table}

\begin{figure}[ht]
  \begin{center}
    \includegraphics[width=0.36\textwidth]{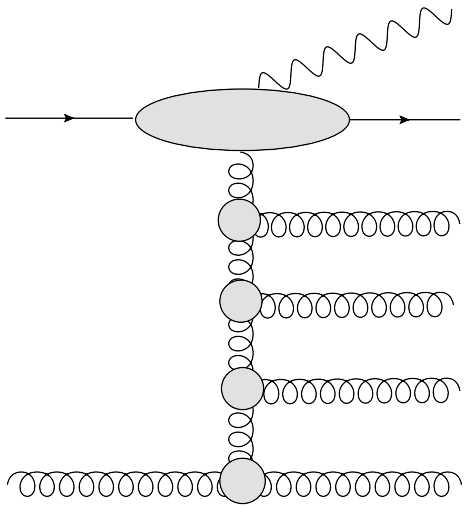}
    \caption{The dominant configuration for $qg \to W + ng$ scattering in
      the high-energy limit. The blobs represent generalised
      vertices. }\label{fig:blob}
  \end{center}
\end{figure}

From Tables~\ref{tab:scalevar} and~\ref{tab:xsecCut} we see that the
ratio $R^\pm(n)$ drops quite significantly from $n=0$ to $n=1$, and
then increases steadily with $n$ thereafter. This can be understood by
considering the dominant subprocess contributions.
Table~\ref{tab:breakdown} shows the subprocess breakdown of $(W^+ +
W^-)+n$~jet production at the 14~TeV LHC. For large $n$ the fractions
appear to stabilise with $Qg \equiv (q + \bar q)g$ production
dominating. This can be understood as arising from the dominance of
`BFKL-like' configurations, in which the scattering amplitudes are
dominated by $t-$channel gluon exchange with the $W^+$ ($W^-$) emitted
off a positively (negatively) charged quark or antiquark line,
\beq
g + q^\pm \to W^\pm + q'^\mp + ng ,
\eeq
as illustrated in Fig.~\ref{fig:blob}. In fact this is the basis of
the `high-energy approximation' for $W+$~jets production developed
first in Ref.~\cite{Andersen:2001ja} and more recently and more
comprehensively in Ref.~\cite{Andersen:2009nu} (see also
Ref.~\cite{jeppeall}). It would be interesting to see how well this
high-energy approximation agrees with the exact results for $R^\pm(n)$
for $n=2,3,4$, since it can easily be extended to higher values of
$n$. Note also that in this approximation the dominant quark
scattering contribution is obtained by replacing the incoming/outgoing
gluon at the bottom of the diagram by a quark or antiquark line (of 5
quark flavours). The effective PDF at the bottom of the diagram is
therefore $g(x,\mu^2) + (4/9)\sum_q [q(x,\mu^2) + \bar q(x,\mu^2)]$.
As the average value of $x$ increases with the number of jets $n$, we
would expect the quark-quark contribution to slowly increase with
respect to the quark-gluon contribution, exactly as seen in
Table~\ref{tab:breakdown}.

\begin{table}[ht]
  \centering
  \begin{tabular}{|l|c|c|c|}
    \hline\hline
    $n$    & $QQ$ & $Qg$ & $gg$ \\ \hline
    0      & 100  &   0  &  0   \\
    1      &  18  &  82  &  0   \\
    2      &  21  &  73  &  6   \\
    3      &  23  &  70  &  7   \\
    4      &  25  &  67  &  8   \\
    \hline\hline
  \end{tabular}
  \caption{Parton subprocess breakdown (in per cent) of leading-order
    $W^\pm +n$~jet production at the LHC ($\sqrt{s} = 14$~TeV), with
    the standard set of cuts and scale choice $\mu_R=\mu_F=\MW$. Here
    $Q = q$~or~$\bar q$.}
  \label{tab:breakdown}
\end{table}

\begin{figure}[ht]
  \begin{center}
    \includegraphics[width=0.5\textwidth]{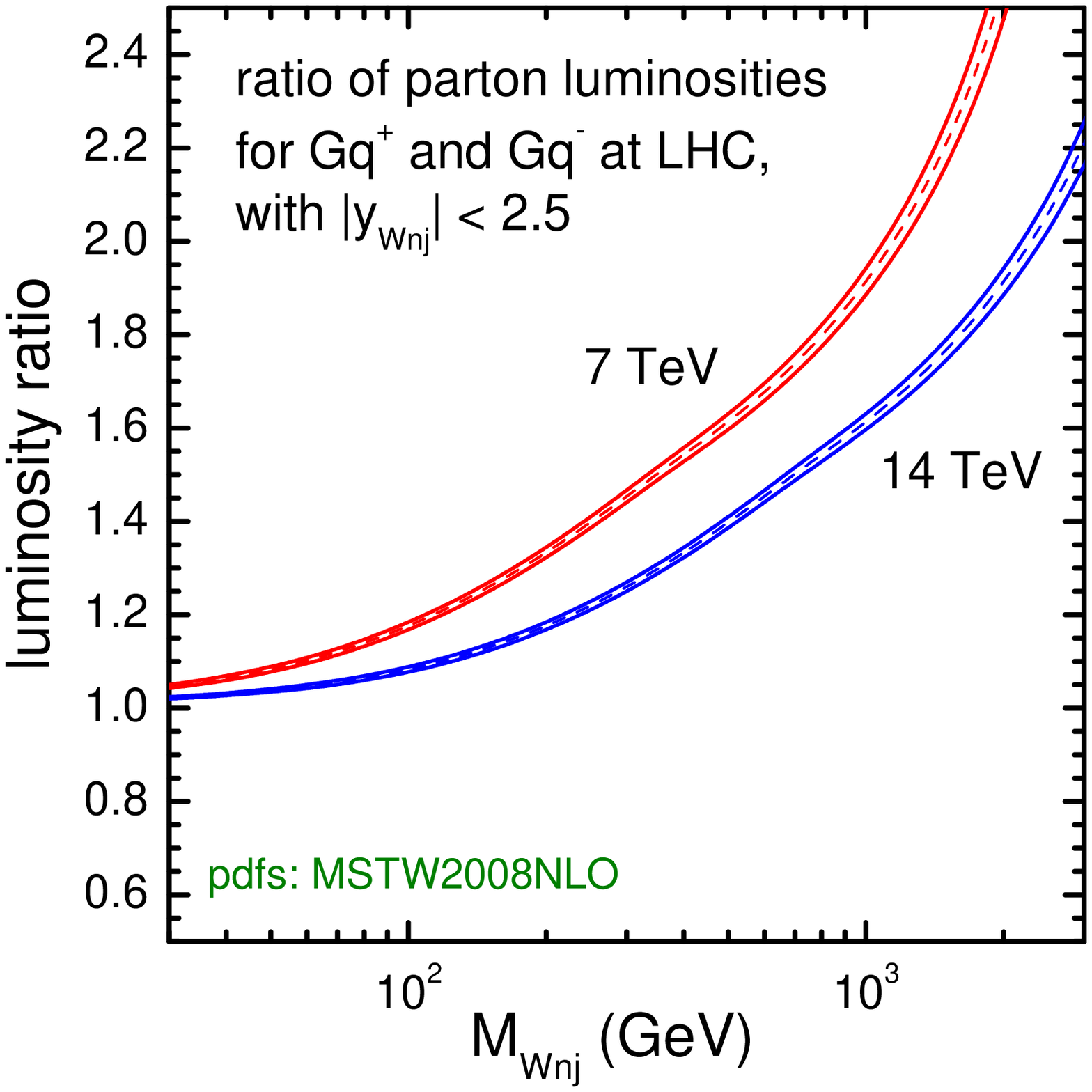}
    \caption{The parton-parton luminosity ratio $\tilde{R}^{\pm}$
      defined in the text, as a function of the subprocess energy
      $\sqrt{\hat{s}} = M_{Wn{\rm j}}$.}
    \label{fig:lumi_qpqm}
  \end{center}
\end{figure}

Because of this $t-$channel dominance, we can relate the $W^\pm
+n$~jet cross-section ratio to the ratio of the parton-parton
differential luminosities
\beq \tilde{R}^\pm \equiv \frac{\partial{\cal
    L}/\partial\hat{s}(q^+ G) }{ \partial{\cal L}/\partial\hat{s}(q^-
  G) }, 
\label{eq:lumirat}
\eeq 
with $q^+ = u + c + \bar d + \bar s $, $q^- = d + s + \bar u + \bar
c$, $G = g + (4/9)\sum_q [q + \bar q ]$, where the sum is over 5 quark
flavours, and $\sqrt{\hat{s}} = M_{Wn{\rm j}}$, the invariant mass of
the $W^\pm + n$ jet system.  Because the final states are restricted
to be central by the cuts imposed, when computing $\tilde{R}^\pm$ we
restrict the rapidity of the $W+n$ jet system to lie within $|y|<2.5$.
The ratio $\tilde{R}^\pm$ is shown as a function of $\sqrt{\hat{s}}$
in Fig.~\ref{fig:lumi_qpqm}. According to the exact calculations, the
average value of $\sqrt{\hat{s}}$ for $n=1,2,3,4$~jet production at
14~TeV is 200, 500, 800, 1300~GeV respectively. As can be seen in
Fig.~\ref{fig:lumi_qpqm}, this corresponds to luminosity ratios in the
range $[1.3, 1.6]$, in line with the exact results.  We can also now
understand why the $R^\pm(n)$ ratio decreases when going from 0 to 1
jets. The leading $q \bar q$ and $qg$ contributions are dominated by
$u\bar q$ ($d\bar q$) and $ug$ ($dg$) scattering for $W^+$ ($W^-$)
production, and hence for these contributions the charge asymmetry
ratio tracks the $u/d$ pdf ratio. However for $n=1$~jet, there is a
non-negligible contribution from $(\bar d + \bar s + c)g \to W^+ \bar
q$ and $(\bar u + s + \bar c)g \to W^- \bar q$ which is approximately
($W$) charge symmetric, and hence serves to reduce $R^\pm(n)$ for $n
\geq 1$. We note also that in $W+1$~jet events, tagging the
final-state jet as a charm-quark jet would give $R^\pm(1,{\rm charm})
\approx 1$, since $s \approx \bar s$ at small $x$.

\begin{figure}[ht]
  \begin{center}
    \includegraphics[width=0.5\textwidth]{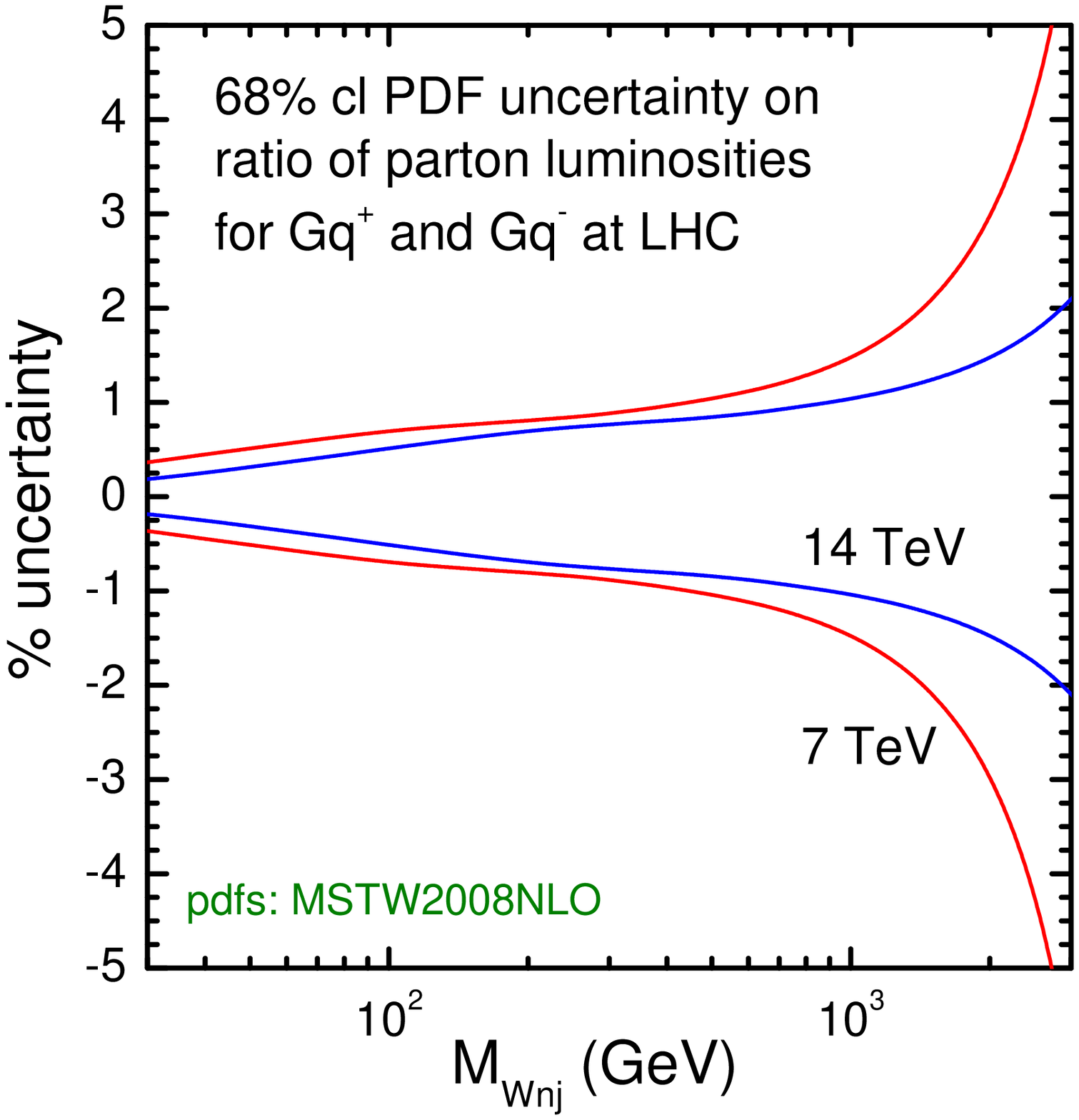}
    \caption{PDF (68\% cl) uncertainty on the parton-parton luminosity
      ratio $\tilde{R}^{\pm}$ defined in Eq.~(\ref{eq:lumirat}), as a function of the subprocess energy
      $\sqrt{\hat{s}} = M_{Wn{\rm j}}$. }
    \label{fig:lumi_qpqm_err}
  \end{center}
\end{figure}

The association of the exact cross-section ratio with the parton
luminosity ratio of Fig.~\ref{fig:lumi_qpqm} enables us to readily
estimate the PDF uncertainty on $R^\pm(n)$.
Fig.~\ref{fig:lumi_qpqm_err} shows the corresponding MSTW2008 NLO PDF
(68\% cl) uncertainty (see Ref.~\cite{Martin09:2009iq} for a full
discussion) on the $q^+G/q^-G$ parton luminosity ratio. Note that as
the subprocess energy increases from 100~GeV to 1~TeV, the uncertainty
increases from approximately 0.5\% to 1\%. This suggests that a
reasonably conservative estimate of the PDF uncertainty on $R^\pm(n)$
for $n \leq 4$ is $\sim\pm 1\%$, in line with the results for the
ratio of total cross sections given in
Table~\ref{tab:w+w-totalrat}. We have checked that this is indeed the
case for the exact ratio calculations up to and including $n=3$ at
leading order.

Putting everything together, we obtain the predictions for $R^\pm(n)$
at the LHC, including scale and PDF uncertainties shown in
Table~\ref{tab:summary}.  Results for both 14~TeV and 7~TeV collider
energy are shown. The ratios are all systematically larger at the lower
collider energy, reflecting the increase in the $u/d$ PDF ratio at
larger $x$ (see also Tables~\ref{tab:w+w-totalrat},\ref{tab:xsecCut}
and Fig.~\ref{fig:lumi_qpqm}). The ratios are calculated exactly at
NLO for $n=0,1,2$ and estimated at LO for $n=3,4$. The central values
are obtained with scale choice $\mu_R=\mu_F=\HT$, and the scale
variation uncertainty is conservatively chosen to encompass all the
predictions for different scales shown in Table~\ref{tab:scalevar}.
Observing the change in $R^\pm(n)$ from LO to NLO for $n=1,2$ and also
for $n=3$ in Ref.~\cite{Berger:2009zg} with a similar set of cuts, we
might expect NLO predictions for $n=3,4$ to decrease slightly from our
LO central values.  Note that restricting the scale variation to
$[\,0.5\HT, 2\HT\,]$ would give a significantly smaller scale
uncertainty. Overall, the combined theoretical uncertainty on the
ratio predictions increases slightly with the number of jets, but is
never more than $\pm 3\%$.

We emphasise again that the PDF uncertainties shown in
Table~\ref{tab:summary} are obtained using the MSTW2008 sets only.  As
shown for example in Fig.~69 of Ref.~\cite{Martin09:2009iq}, other
available PDF sets give central predictions for the charge asymmetry
ratio that are slightly offset from the MSTW2008 predictions, but with
similar uncertainties.\footnote{For example, using
  CTEQ6.6(NLO)~\cite{Nadolsky:2008zw} instead of MSTW2008(NLO) pdfs
  gives charge asymmetry ratios that are approximately $2-3\%$ larger
  for $n=1-4$.}  The reasons for this are only partially understood,
see for example the discussion in the PDF4LHC workshop series
\cite{PDF4LHC}.

\begin{table}[ht]
  \centering
  \begin{tabular}{|l|c|c|}
    \hline\hline
    $n$& $\sqrt{s} = 7$~TeV & $\sqrt{s} = 14$~TeV \\ \hline
    0  & $1.52\;\pm\;0.01\;({\rm scl})\;\pm\;0.02\;({\rm pdf})$& $1.31\;\pm\;0.01\;({\rm scl})\;\pm\;0.01\;({\rm pdf})$ \\
    1  & $1.45\;\pm\;0.01\;({\rm scl})\;\pm\;0.01\;({\rm pdf})$& $1.27\;\pm\;0.01\;({\rm scl})\;\pm\;0.01\;({\rm pdf})$ \\
    2  & $1.56\;\pm\;0.02\;({\rm scl})\;\pm\;0.02\;({\rm pdf})$& $1.33\;\pm\;0.02\;({\rm scl})\;\pm\;0.01\;({\rm pdf})$ \\
    3  & $1.72\;\pm\;0.03\;({\rm scl})\;\pm\;0.03\;({\rm pdf})$& $1.45\;\pm\;0.03\;({\rm scl})\;\pm\;0.02\;({\rm pdf})$ \\
    4  & $1.87\;\pm\;0.04\;({\rm scl})\;\pm\;0.03\;({\rm pdf})$& $1.55\;\pm\;0.04\;({\rm scl})\;\pm\;0.02\;({\rm pdf})$ \\
\hline\hline
  \end{tabular}
  \caption{Predictions for $R^\pm(n)$ at the LHC for the standard set
    of cuts defined in Eq.~(\ref{eq:cutsdef}).  The ratios are
    calculated exactly at NLO for $n=0,1,2$ and estimated at LO for
    $n=3,4$. The central values are obtained with scale choice
    $\mu_R=\mu_F=\HT$.  The scale (scl) and PDF uncertainties are
    displayed separately.}
  \label{tab:summary}
\end{table}

\begin{figure}[ht]
  \begin{center}
    \includegraphics[width=0.57\textwidth]{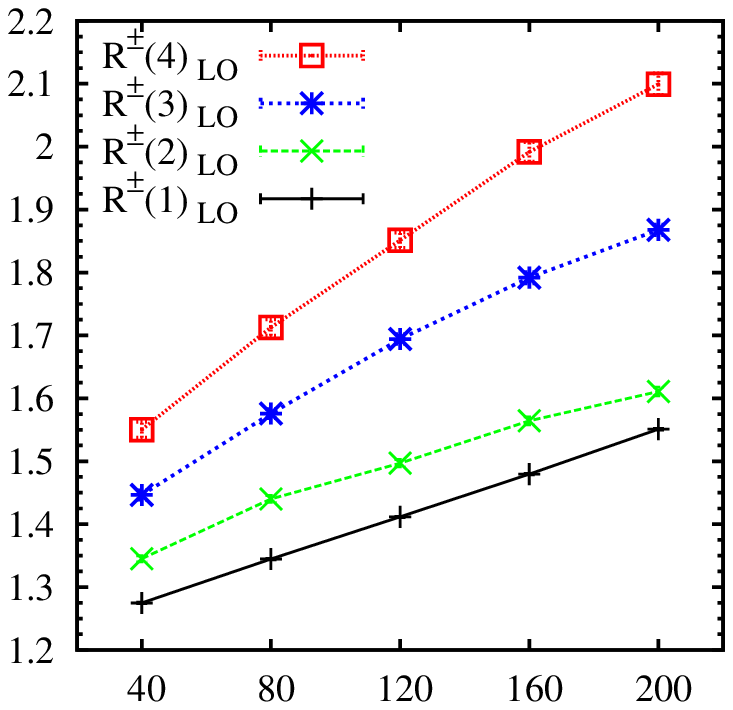}
    \put(-290,110){\rotatebox{90}{$R^{\pm}$}}
    \put(-170,-15){$p_{Tj}^{\rm min}$ (GeV)}
    \put(-80,-20){$ $}
    \caption{Variation of $R^\pm(n)$ as a function of $p_{Tj}^{\rm
        min}$.  The ratios are obtained with scale choice
      $\mu_R=\mu_F=\HT$.  For simplicity, the calculations are
      performed at leading order. }
    \label{fig:RvsPtj}
  \end{center}
\end{figure}

Finally, we show in Fig.~\ref{fig:RvsPtj} the values of $R^\pm(n)$ for
different values of $p_{Tj}^{\rm min}$.  We see that the ratios
increase with $p_{Tj}^{\rm min}$, again a reflection of the increase
in the $u/d$ PDF ratio at larger $x$. This makes the SM charge
asymmetry more prominent for hard, multijet final states, which could
assist the detection of New Physics contributions.

\section{Conclusions}

In this paper, we have studied the cross-section ratios, $R^\pm(n)$,
of $\Wp + {\rm jets}$ and $\Wm + {\rm jets}$ production for different
number of jets $n$ at the LHC.  We exploited the charge asymmetry
nature of proton-proton collisions at the LHC to demonstrate that it
provides an additional handle to study New Physics signals in the
$W(\to l\nu) + {\rm jets}$ channel, where typically the charged
leptons are produced in equal quantities, and therefore any deviation
of $R^\pm$ from the SM expectation would indicate the presence of New
Physics.

Quantitatively, we showed that the $R^\pm$ ratios are remarkably
stable with respect to theoretical uncertainties from scale choices,
higher-order corrections and from PDFs.  We have also demonstrated a
connection between the cross-section ratios $R^\pm(n)$ and the parton
luminosity ratio $\tilde{R}^\pm$ based on arguments from BFKL
dominance, and showed that $R^\pm(n)$ can be reasonably well
approximated by $\tilde{R}^\pm$ with a suitable choice of subprocess
invariant mass.

Given the simple and robust nature of this observable, it could prove
useful for studying New Physics phenomena, particularly at the early
stages of LHC operation.  In a future study we will consider specific
NP scenarios for which the method could be applicable.

\section*{Acknowledgements}
We thank Bobby Acharya for useful discussions and John Hill for suggestions on various computing issues.

\end{document}